\documentclass[reprint,amsmath,amssymb,aps,prl,]{revtex4-2}
\usepackage[utf8]{inputenc}
\usepackage{graphics, bm, psfrag, amsmath, amssymb, epsfig, grffile, float, bbold}
\usepackage{dsfont}
\usepackage{color}
\usepackage{hyperref}
\usepackage[shortlabels]{enumitem}
\usepackage{amsfonts,amsthm,mathtools}
\usepackage{latexsym,graphicx}
\usepackage{amscd}
\usepackage{natbib} 
\usepackage{xcolor}
\usepackage{soul}
\usepackage[capitalize]{cleveref}
\usepackage[percent]{overpic}

\newcommand{\ee}{{\rm e}}

\newcommand{\bk}{\mathbf{k}}

\newcommand{\DF}{\Delta_{\mathrm{F}}}
\newcommand{\phiF}{\phi_{\mathrm{F}}}

\newcommand{\phiNM}{\phi_{\mathrm{NM}}}

\newcommand{\phiAM}{\phi_{\mathrm{AM}}}
\newcommand{\DAM}{\Delta_{\mathrm{AM}}}

\newcommand{\LnT}{\mathrm{Ln}_T}

\newcommand{\eps}{\epsilon}

\newcommand{\cF}{\mathcal{F}}

\newcommand{\HHF}{H_{\mathrm HF}}

\newcommand{\pdag}{{\phantom\dag}} 

\newcommand{\Tr}{\mathrm{Tr}}

\newcommand{\xx}{w}

\begin{document}

\title{Mixed phases in a Fermi--Hubbard model describing  altermagnetism}

\author{E. Langmann$^{1}$ and J. Lenells$^{2}$}

\address{$^1$Department of Physics, KTH Royal Institute of Technology, 106 91 Stockholm, Sweden
	\\
	$^{2}$Department of Mathematics, KTH Royal Institute of Technology, 100 44 Stockholm, Sweden
}
\email{langmann@kth.se}
\email{jlenells@kth.se}


\begin{abstract}
We study an extension of the 2D Fermi--Hubbard model, which was recently introduced in [Das et al., Phys. Rev. Lett. \textbf{132}, 263402 (2024)] and shown to describe altermagnetism that can be studied in cold atom systems. 
Using an updated Hartree--Fock method that can detect instabilities towards phase separation, we show that the model is in a mixed phase in large parts of the parameter regime at half-filling. We argue that the occurrence of a mixed phase is an indication of exotic physics which, in this model, occurs in parameter regimes accessible in cold atom experiments. 
\end{abstract}

\maketitle

\paragraph{Introduction.---}In a recent paper, Das--Leeb--Knolle--Knap (DLKK) introduced an interesting two-parameter generalization of the 2D (Fermi) Hubbard model which, as they showed, can describe  altermagnetism at half-filling \cite{D2024}. They also showed that this model can be realized in cold atom systems, and that one characteristic of altermagnetism, the anisotropic spin transport, can be probed with trap-expansion experiments \cite{D2024}.

To obtain their results, DLKK \cite{D2024} used  restricted Hartree--Fock (HF) theory with a famous ansatz employed by Penn in his classic work on the mean-field phase diagram of the 3D Hubbard model from 1966 \cite{P1966}.  This method is standard but, as emphasized by Penn \cite[Sec. III]{P1966}, it is important to check the stability of mean-field solutions to avoid wrong results. Known methods to test stability at the time, and used up to this day, only make it possible to investigate {\em local} stability, or stability within a restricted class of HF states. Today, it is known from numeric studies that, in the 2D Hubbard model away from half-filling, simple translation-invariant HF states can be unstable towards states where translational invariance is broken in complicated ways \cite{ZG1989,PR1989,V1991}; such  instablities can be detected in a simpler way as a mixed phase \cite{LW1997,LW2007}, and they indicate exotic physics as observed in, for example, high-temperature superconductors described by a 2D Hubbard-like model \cite{A1987}. 
However, there is a general belief that such instabilities only occur away from half-filling. 

In this paper, we show that in the DLWW model \cite{D2024}, mixed phases are ubiquitous even at half-filling. By testing the stability of the HF solutions presented in \cite{D2024}, we found that some of these solutions are unstable and therefore should be discarded. For other parameters, we could confirm results in Ref.~\cite{D2024} but, as we show, mixed phases at half-filling occur in large parts of the phase diagram of the DLKK model.
Our main result is that \emph{the DLKK model can bring exotic physics relevant for high-temperature superconductivity into reach of experiments on cold atom systems}; see Figs.~\ref{fig2}--\ref{fig4}.

HF theory is a basic method, and DLKK \cite{D2024} used it as taught in textbooks, as many papers before. But this simplified method is not reliable: as our results show, it can lead to results that are qualitatively wrong. Thus,  HF theory is in need of an update; in an  accompanying detailed paper, we present such an update, and we correct some well-known but incorrect results about the Hubbard model in the literature \cite{LL2024A}. 

\paragraph{DLKK model.---}
The model introduced by DLKK \cite{D2024} is defined by the Hamiltonian 
\begin{equation}\label{H} 
	H=\sum_{i,j}\sum_{\sigma} \big( t_{i,j}^\pdag-\mu\delta_{i,j}\big) c^\dag_{i,\sigma} c^\pdag_{j,\sigma} 
	 + \sum_i Un_{i,\uparrow}n_{i,\downarrow}
\end{equation}
with fermion operators $c^{(\dag)}_{i,\sigma}$ labeled by lattice sites $i$ on a square lattice,  $i=(i_1,i_2)$ with $i_a=1,\ldots,L$ for $a=1,2$ and periodic boundary conditions, and $\sigma =\; \uparrow,\downarrow$ a spin index; $n_{i,\sigma}= c_{i,\sigma}^\dag c_{i,\sigma}^\pdag$ are density operators (with eigenvalues 0 and 1), $U>0$ is the coupling parameter, $\mu$ is the chemical potential, and the hopping matrix $t_{i,j}$ is $-t$ if $j=i\pm e_a$ for $a=1,2$ (with $e_1=(1,0)$ and $e_2=(0,1)$), $-t'(1+\delta(-1)^{i_1+i_2}))$ for $j=i\pm (e_1+e_2)$, $-t'(1-\delta(-1)^{i_1+i_2}))$ for $j=i\pm (e_1-e_2)$, and 0 otherwise (see \cite[Fig.~1(a)]{D2024} for a visualization of $t_{i,j}$). 
Thus, the model depends on three parameters: $U/t>0$, $t'/t$ and $\delta>0$, with the special case $t'=0$ corresponding to the usual 2D Hubbard model. 
We work with the grand canonical ensemble and determine the chemical potential $\mu$ by the density $\rho = (1/L^2)\sum_i\langle n_{i,\uparrow}+n_{i,\downarrow}\rangle $; 
the case of main interest to us is half-filling $\rho=1$ but, as will be explained, it is important to allow also for other density values to be able to detect possible instabilities. As usual, we use a finite  lattice with $L^2$ sites and take the thermodynamic limit $L\to\infty$ at the end of computations. 

\paragraph{Hartree--Fock theory.---}As is well-known (but often ignored), HF theory is a variational method where one tries to find the best possible approximation of the Gibbs state at temperature $T>0$ by a quasi-free thermal state given a Hamiltonian depending on real-valued variational fields $\phi_i$ and $B_i$ as follows ($H_0=H|_{U=0}$ is the non-interacting part of the Hubbard Hamiltonian $H$ in \eqref{H})
\begin{equation}\label{HHF} 
	\HHF = H_0 + \sum_i\big( \phi_i n_i + B_iS_i), 	
\end{equation} 
where $n_i=n_{i,\uparrow}+n_{i,\downarrow}$ is the local density operator and $S_i=n_{i,\uparrow}-n_{i,\downarrow}$ is the spin operator projected to the $z$-direction (as in \cite{D2024}, we simplify our discussion by restricting to states where spin rotation invariance can only be broken in the $z$-direction; see e.g.\ \cite{V1991,BLS1994,LW2007} for the general case). To obtain HF equations from the Hamiltonian in \eqref{H}, one inserts $n_{i,\sigma}=\langle n_{i,\sigma}\rangle + \delta n_{i,\sigma}$ into the interaction and ignores terms which are quadratic in the fluctuations $\delta n_{i,\sigma}$: 
$
n_{i,\uparrow}n_{i,\downarrow} \approx 
\langle n_{i,\uparrow}\rangle n_{i,\downarrow}  + n_{i,\uparrow} \langle n_{i,\downarrow}\rangle  - \langle n_{i,\uparrow}\rangle\langle n_{i,\downarrow}\rangle  
$; 
this leads to $H\approx \HHF-E_0$ where 
\begin{equation*} 
\HHF = H_0 + \frac{U}{2}\sum_i \big(\langle n_i\rangle n_i-\langle S_i\rangle S_i \big) 	
\end{equation*} 
and $E_0= (U/4)\sum_i\big( \langle n_i\rangle \langle n_i \rangle -\langle S_i\rangle \langle S_i\rangle \big)$. By comparing with \eqref{HHF}, one obtains 
\begin{equation}\label{HFeqs}  
\phi_i = \frac{U}{2}\langle n_i\rangle,\quad 
B_i = -\frac{U}{2}\langle S_i\rangle,	
\end{equation} 
which become the HF equations determining the variational fields if the expectation values are interpreted as the thermal state of the HF Hamiltonian in \eqref{HHF}:  
 $\langle A\rangle = \Tr(A\ee^{-\HHF/T})/\Tr(\ee^{-\HHF/T})$ for $A=n_i$, $S_i$ (we set $k_B=\hbar=1$).

Following prominent previous work on special cases of their model \cite{H1985,LH1987}, DLKK \cite{D2024} restricted attention to HF states of the form
\begin{equation}\label{ansatz} 
	\phi_i = \phi,\quad B_i=\Delta(-1)^{i_1+i_2},
\end{equation}  
and studied two different solutions of the HF equations: (i) $(\phi,\Delta)=(\phiAM,\DAM)$ with $\DAM>0$ describing an altermagnetic (AM) state (which reduces to an anti-ferromagnetic state for $\delta=0$), (ii) $(\phi,\Delta)=(\phiNM,0)$ describing a normal metal (NM) state; they also distinguished metallic and insulating AM states referred to as AMM and AMI, respectively \cite{D2024}. As observed by DLKK \cite{D2024},  for a solution of the HF equations of the form \eqref{ansatz}, $\langle n_i\rangle$ is $i$-independent and equal to the average density: $\rho=2\phi/U$, and for many parameter values one can find a unique chemical potential $\mu$ allowing for an AM solution such that $2\phiAM/U=1$; they proposed that such a solution describes an AM state of the system at half-filling. However, this proposal has  an implicit assumption: it assumes that the AM solution is stable, i.e., that (i) its free energy $\cF_{\mathrm{AM}}$ is a local minimum (local stability) and (ii) this free energy $\cF_{\mathrm{AM}}$ is lower than the free energy $\cF_{\mathrm{NM}}$ of the NM solution at the same value of $\mu$ (global stability within the HF ansatz). If any of these two conditions is not fulfilled, then this AM solution cannot be realized in the system and should be discarded (since this point is often ignored in HF studies in the literature, we justify it in more detail in an appendix). For this reason, we propose below  a method which allows us to remain open to the possibility that solutions of HF equations are unstable. 

In the limit $L\to\infty$, the free energy (density) of a HF state described by the ansatz in \eqref{ansatz} is given by 
\begin{equation}\label{cFansatz}  
\cF = \frac{\Delta^2-\phi^2}{U}
-\frac12\sum_{r,r'=\pm}\int \frac{d^2\bk}{(2\pi)^2}\LnT(E_{r,r'}(\bk)) 	
\end{equation} 
with $\LnT(E)=T\ln(1+\exp(-E/T))$, the integral over the Brillouin zone: $\bk=(k_1,k_2)$ with $-\pi\leq k_a\leq \pi$ for $a=1,2$, and the following AM band relations \cite{D2024}:
\begin{equation} 
E_{r,r'}(\bk)  = \eps_1+\phi-\mu+r\sqrt{\eps_0^2+(\eps_2+r'\Delta)^2}, 
\end{equation} 
$\eps_0=-2t[\cos(k_1)+\cos(k_2)]$, $\eps_1=-4t'\cos(k_1)\cos(k_2)$, $\eps_2=-4t'\delta \sin(k_1)\sin(k_2)$ (this is well-known in special cases \cite{BLS1994,LW2007} and, by using results in \cite{D2024}, it straightforwardly generalizes to the present case \cite{LL2024A}). 
Moreover, this function $\cF=\cF(\phi,\Delta;\mu)$ gives the free energies of the AM and NM states as $\cF_{\mathrm{AF}}(\mu)=\cF(\phiAM,\DAM;\mu)$ 
and $\cF_{\mathrm{NM}}(\mu)=\cF(\phiNM,0;\mu)$,  with  $(\phi,\Delta)=(\phiAM,\DAM)$ and $(\phiNM,0)$ the solutions of the HF equations obtained with the ansatz in \eqref{ansatz}, as discussed above. It is important to note that the corresponding densities are given by $\rho_X=-\partial\cF_X/\partial\mu$ for $X= \mathrm{AM}, \mathrm{NM}$: at the same value of $\mu$, the AM and NM solutions correspond to different densities in general.

\begin{figure}
\vspace{0.4cm}
\begin{center}
\hspace{-.6cm}
\begin{overpic}[width=.46\textwidth]{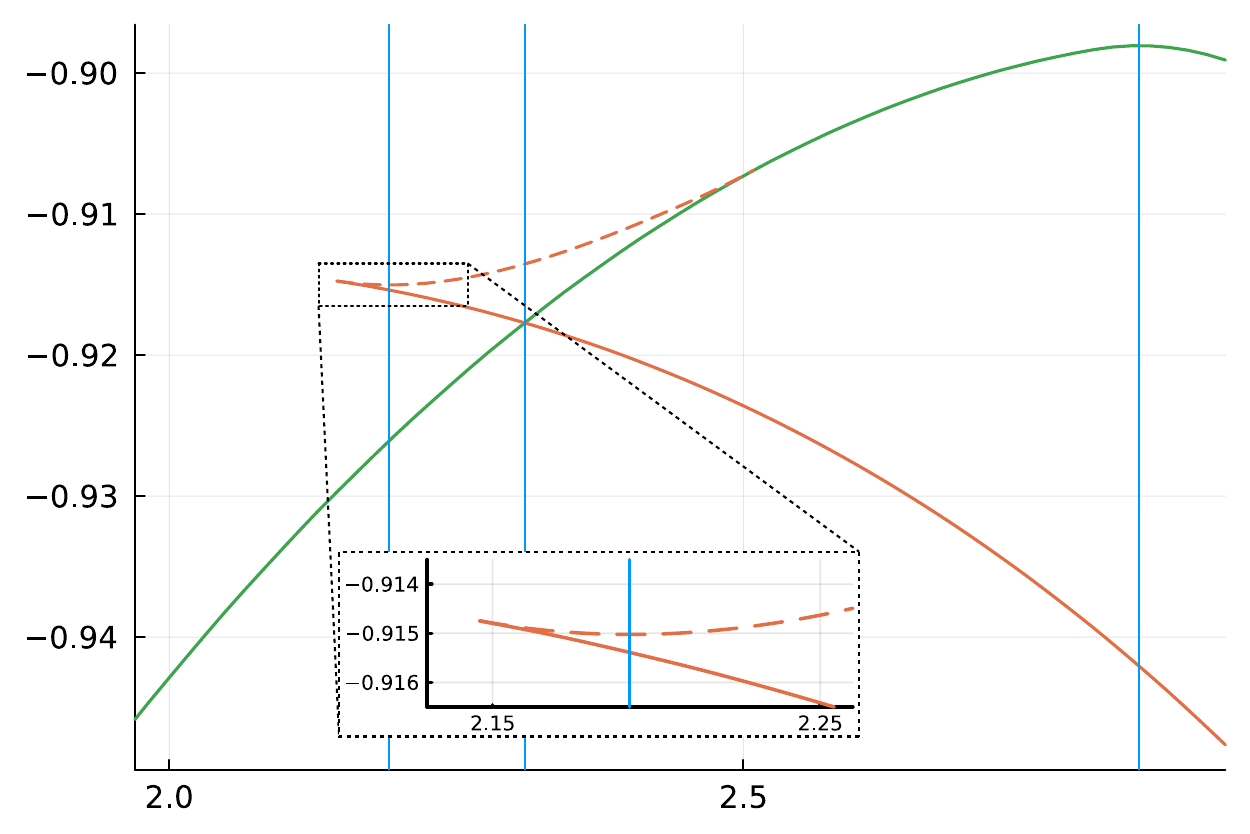}
      \put(5.7,66.5){\footnotesize $\cF + \mu$}
      \put(100,4.5){\footnotesize $\mu$}
      \put(89.8,2.6){\footnotesize $\mu_{\mathrm{NM}}$}
      \put(29.9,2.6){\footnotesize $\mu_1$}
      \put(41,2.6){\footnotesize $\mu_c$}
      \put(49,8.7){\tiny $\mu_1$}
   \end{overpic}
\caption{\label{fig1} HF free energies of the DLKK model for $U/t=3.5$, $t’/t=0.3$, $\delta=0.9$, and $T=0$, setting $t=1$: shown are $\cF+\mu$, with $\cF$ the free energy and $\mu$ the chemical potential, as a function of $\mu$ for the altermagnetic (AM: red curves) and normal metal (NM: green curve) solutions, as discussed in the main text. Inset: Zoom-in of the AM solution (red dashed) used for \cite[Fig.~1(b)]{D2024}.}
     \end{center}
\vspace{-0.4cm}
\end{figure}
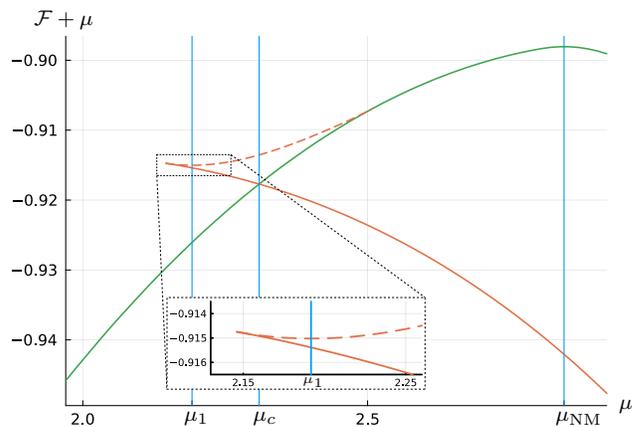

To solve the HF equations and, at the same time, check for stability, we use the following procedure: (i) Fix $\mu$, (ii) find {\em all} solutions $(\phi,\Delta)$ of the HF equations in \eqref{HFeqs}, (iii) compute the free energy $\cF$ in \eqref{cFansatz} for each solution. In Fig.~\ref{fig1}, we present such results for the parameters used by DLKK in \cite[Fig.~1(b)]{D2024}: $t'/t=0.3$, $\delta=0.9$, $U/t=3.5$, $T=0$;  
we plot $\cF_X+\mu$ vs.\ $\mu$ to make half-filling $\rho_X=1$ clearly visible as  $\partial(\cF_X+\mu)/\partial\mu=0$ (horizontal tangent).
For $\mu<\mu_{\mathrm{min}}\approx 2.146t$, there is no AM solution with $\Delta\neq 0$, for  $\mu_{\mathrm{min}}<\mu<\mu_{\mathrm{max}}\approx 2.511t$, there are two AM solutions (solid and dashed red curves), and for $\mu>\mu_{\mathrm{max}}$ there is one  (solid red); the NM solution with $\Delta=0$ (green) exists for all $\mu$. 
Clearly, there is a single AM solution (dashed red) with $\rho_{\mathrm{AM}}=1$ at $\mu=\mu_{1}\approx 2.192t$; see inset in Fig.~\ref{fig1}.  This is the solution which DLKK used to compute  \cite[Fig.~1(b)]{D2024}: to check this, we computed the effective bands for this solution, and we found perfect agreement with \cite[Fig.~1(b)]{D2024}. But, as is clear from Fig.~\ref{fig1}, this AM solution is not stable and thus cannot be realized in the system: both the other AM solution (solid red) and the NM solution have smaller free energy at $\mu=\mu_1$ than this solution. Thus, \cite[Fig.~1(b)]{D2024} needs to be corrected. 
Furthermore, as is clear from Fig.~\ref{fig1}, there is a single NM solution with $\rho_{\mathrm{NM}}=1$ at $\mu=\mu_{\mathrm{NM}}\approx 2.845t$, but this half-filled solution is unstable since, at this value of $\mu$, there is an AM solution (solid red) with smaller free energy.
We thus conclude: {\em for these parameters, there is no stable solution at half-filling such that the ansatz in \eqref{ansatz} holds true  throughout the system}. As we now explain, this result proves that there exists a half-filled variational state with lower free energy than both the half-filled AM and NM states, and this state describes a mixed phase. 

\paragraph{Mixed phase.---}By writing the ansatz \eqref{ansatz} and computing the free energy in \eqref{cFansatz}, we actually know the free energy density for a plethora of variational HF  states where macroscopic regions in AM and NM phases are mixed \cite{LW1997}. These states are characterized not only by the parameters $\phiNM$, $\phiAM$ and $\DAM$ but also by a parameter $\xx$ in the range $0\leq \xx\leq 1$; $\xx$ is the fraction of the system in the AM phase, with the limiting cases $\xx=0$ and $\xx=1$ corresponding to the pure NM and AM phases, respectively.  
 
To construct such a variational state, we divide the spatial lattice in two disjoint regions $V_0$ and $V_1$,  with $N_0$ and $N_1=L^2-N_{0}$ the number of sites in $V_0$ and $V_1$, respectively; for examples, one could choose positive intergers $\ell_a\leq L$ for $a=1,2$  and take as $V_0$ all sites $i=(i_1,i_2)$ with $i_a=1,\ldots,\ell_a$ so that $N_0=\ell_1\ell_2$ but, clearly, there are many other ways to construct such a state. We now make the following ansatz for the HF Hamiltonian in \eqref{HHF}: for $i\in V_0$, $\phi_i$ and $B_i$ are as in \eqref{ansatz} with $(\phi,\Delta)=(\phiAM,\DAM)$, and for $i\in V_1$, $\phi_i$ and $B_i$ are as in \eqref{ansatz} with $(\phi,\Delta)=(\phiNM,0)$. Clearly, this ansatz describes a state where fractions $N_{0}/L^2\equiv \xx$ and $N_{1}/L^2=1-\xx$ of the lattice sites are in the AM and NM phases, respectively. If we take the thermodynamic limit such that $\xx$ remains fixed and the pure-phase regions become macroscopically large, then the free energy density for such a mixed HF state can be computed by the following argument: the contribution of each site in the AM regions to the total free energy is the pure-phase value, $\cF_{\mathrm{AF}}$, up to corrections that vanish as the distance to a phase boundary becomes large (and similarly for lattice points in NM regions). Since the number of points in the AM and NM phase are $N_0$ and $N_1$, respectively, the total free energy in this state is $F=N_0\cF_{\mathrm{AF}}+N_1\cF_{\mathrm{NM}}+\Delta F$, where $\Delta F$ is the sum of all corrections taking into account the presence of phase boundaties. For large $L$, $\Delta F$ should scale like the total length of the phase boundary; since the latter is proportional to $L$ for large $L$, we conclude that $\lim_{L\to \infty} \Delta F/L^2=0$. Thus, phase boundaries do not affect the free energy density for such mixed states in the thermodynamic limit: as $L\to\infty$, the free energy density $F/L^2$ of the mixed state becomes  
\begin{equation}\label{cFM} 
	\cF_{\mathrm{M}}(\mu) = \xx\cF_{\mathrm{AF}}(\mu) + (1-\xx)\cF_{\mathrm{NM}}(\mu). 	
\end{equation} 
By a similar argument, the average density in such a mixed state is 
\begin{equation}\label{rhoxx}  
\rho=\xx\rho_{\mathrm{AM}}+ (1-\xx)\rho_{\mathrm{NM}};
\end{equation} 
note that this is consistent with  $\rho=-\partial\cF_{\mathrm{M}}/\partial\mu$, as expected. 

Thus, it is clear how to reach half-filling if the AF and NM free energies at half-filling both are unstable as in the example in Fig.~\ref{fig1}: fix the chemical potential $\mu$ at the particular value $\mu_c \approx 2.310$ where the AM and NM free energies intersect: $\cF_{\mathrm{AF}}(\mu_c) = \cF_{\mathrm{NM}}(\mu_c)$; at this value, the AM and NM phases can co-exist, and it is possible to have mixed states with  lower free energy than both the half-filled AM and NM states (see Fig.~\ref{fig1}). Moreover, the parameter $\xx$ characterizing such a mixed state is determined by \eqref{rhoxx} and the condition $\rho=1$ as follows, 
\begin{equation} 
\xx =
\frac{1-\rho_{\mathrm{NM}}}{\rho_{\mathrm{AF}}-\rho_\mathrm{NM}}
\end{equation} 
with $\rho_X=-\partial\cF_X/\partial\mu|_{\mu=\mu_c}$ for $X= \mathrm{AM}, \mathrm{NM}$. In our example in Fig.~\ref{fig1}, we have $\rho_{\mathrm{NM}}\approx 0.935$,  $\rho_{\mathrm{AF}}\approx 1.024$, and thus $\xx\approx 0.731$. 

We stress that finding a mixed phase does not necessarily mean phase separation but only that the states minimizing the HF free energy break translation invariance in complicated ways; the details of these more exotic states can be studied  by more advanced methods (like unrestricted HF theory \cite{V1991}). What our improved  HF method provides, with certainty and by simple means, are the regions in parameter space where such more exotic solutions arise. 

\paragraph{Further results.---} As discussed, Fig.~\ref{fig1} proves that the band relations plotted in \cite[Fig.~1(b)]{D2024} cannot exist in a real system and thus should be corrected; examples of parameters where AM solutions are stable can be obtained from our Fig.~\ref{fig2}. We also checked the HF solutions underlying other  results in \cite{D2024}: we found that there is a mixed phase in some $U$-interval making the phase diagram in \cite[Fig.~2(a)]{D2024} more complicated, but this mixed region is very narrow. Moreover, at higher temperature, this mixed phase disappears: \cite[Fig.~2(b)]{D2024} is correct as it stands. 
Thus, we checked and can confirm that the main results in \cite{D2024} are correct.

\begin{figure}
\vspace{0.4cm}
\begin{center}
\hspace{-.6cm}
\begin{overpic}[width=.4\textwidth]{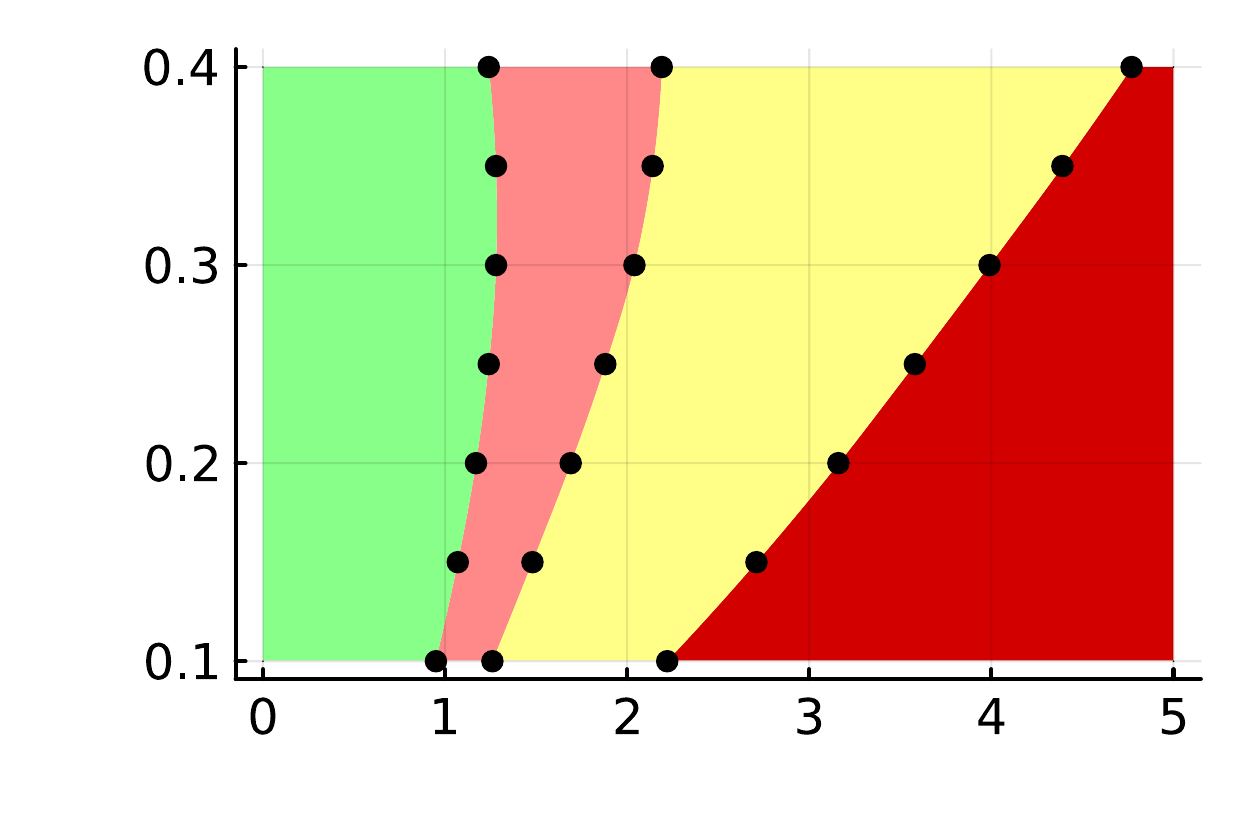}
      \put(16,65){\footnotesize $t'/t$}
      \put(98.5,11){\footnotesize $U/t$}
      \put(26,40){\footnotesize NM}
      \put(40.3,48){\footnotesize AMM}
      \put(57,40){\footnotesize Mixed}
      \put(79,29){\footnotesize AMI}
   \end{overpic}
   \vspace{-0.6cm}
\caption{\label{fig2} Mean-field phase diagram of the half-filled DLKK model for temperature $T=0$ and $\delta=0.9$, varying $U/t$ and $t'/t$. There exists a signifiant mixed phase (yellow) in addition to the NM, AMM and AMI phases found in \cite{D2024}.}   
     \end{center}
\end{figure}

\begin{figure}
\begin{center}
   \vspace{-0.1cm}
\hspace{-.6cm}
\begin{overpic}[width=.4\textwidth]{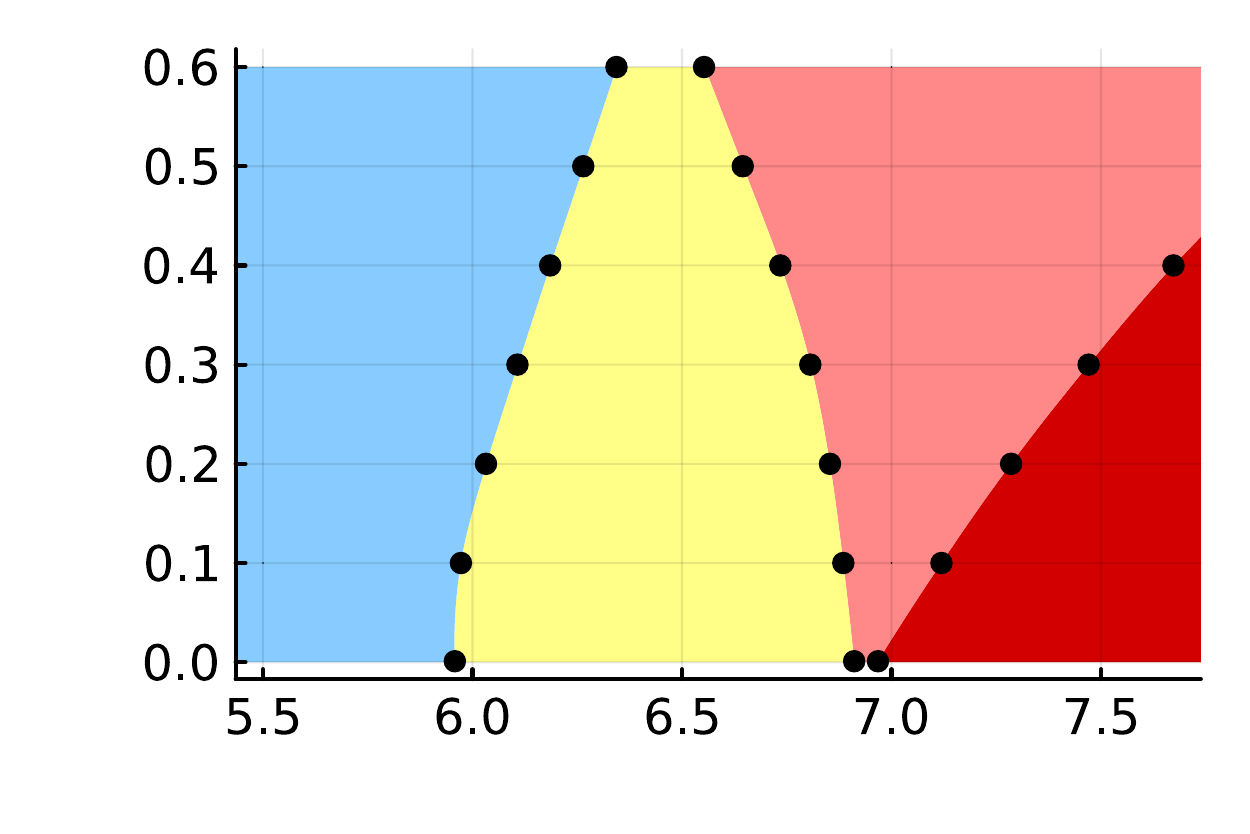}
      \put(15,65.5){\footnotesize $T/t$}
      \put(98.5,11){\footnotesize $U/t$}
      \put(29,32){\footnotesize F}
      \put(48,32){\footnotesize Mixed}
      \put(71,40){\footnotesize AMM}
       \put(83,21){\footnotesize AMI}
  \end{overpic}
   \vspace{-0.6cm}
\caption{\label{fig3} Mean-field phase diagram of the half-filled DLKK model for $t'/t=1.0$ and $\delta=0.2$, varying $U/t$ and $T/t$. There exist significant ferromagnetic (F: blue) and mixed (yellow) phases in addition to the AMM and AMI phases found in \cite{D2024}. 
}
\end{center}
\vspace{-0.4cm}
\end{figure}

\begin{figure}
\begin{center}
\hspace{-1.1cm}
\begin{overpic}[width=.49\textwidth]{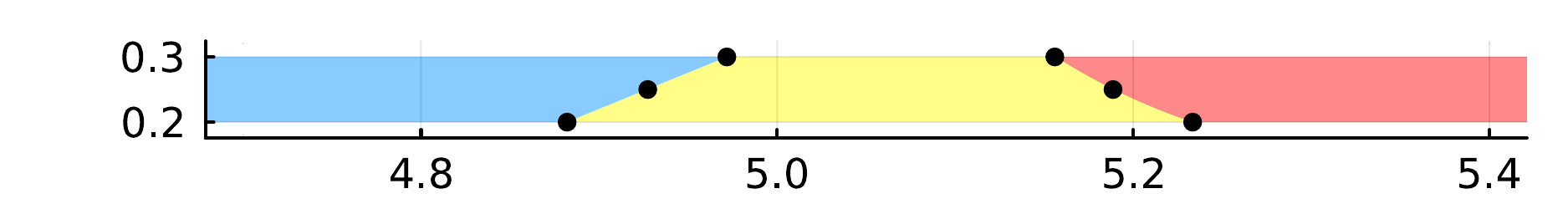}
      \put(40,12){\small $(a)$ $t'=0.8$, $\delta = 0.3$}
      \put(10,12){\footnotesize $T/t$}
      \put(99,3.5){\footnotesize $U/t$}
      \put(26,6.5){\footnotesize F}
      \put(51,6.5){\footnotesize Mixed}
      \put(82,6.5){\footnotesize AMM}
   \end{overpic}
\vspace{.3cm}
	\\
\hspace{-1.1cm}
\begin{overpic}[width=.49\textwidth]{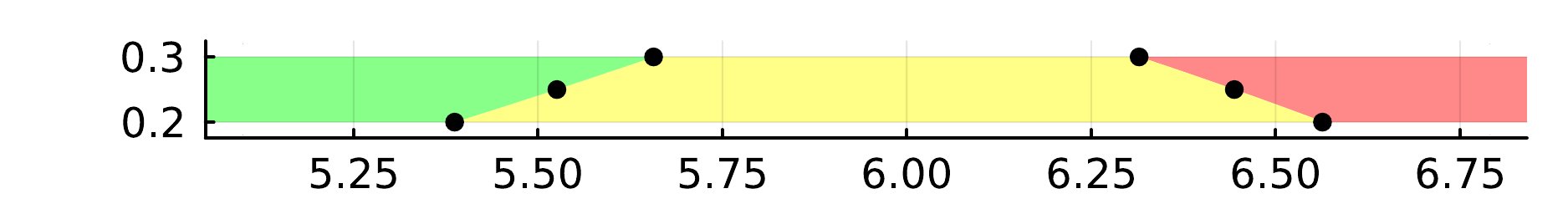}
      \put(40,12){\small $(b)$ $t'=0.6$, $\delta = 1.0$}
      \put(10,12){\footnotesize $T/t$}
      \put(99,3.5){\footnotesize $U/t$}
      \put(19,6.5){\footnotesize NM}
      \put(52,6.5){\footnotesize Mixed}
      \put(86,6.5){\footnotesize AMM}
   \end{overpic}
\vspace{.3cm}
	\\
\hspace{-1.1cm}
\begin{overpic}[width=.49\textwidth]{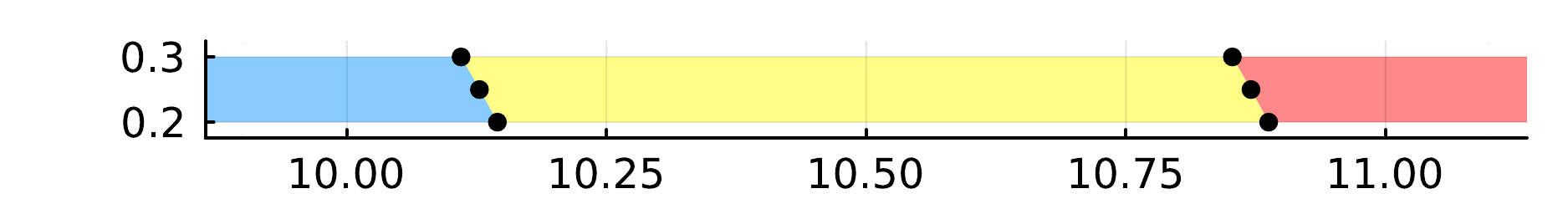}
      \put(40,12){\small $(c)$ $t'=1.5$, $\delta = 0.1$}
      \put(10,12){\footnotesize $T/t$}
      \put(99,3.5){\footnotesize $U/t$}
      \put(21,6.5){\footnotesize F}
      \put(50,6.5){\footnotesize Mixed}
      \put(84,6.5){\footnotesize AMM}
   \end{overpic}
   \vspace{-0.3cm}
   \caption{\label{fig4} Examples of mean-field phase diagrams of the half-filled DLKK model, demonstrating that mixed phases (yellow regions) are ubiquitous even at larger temperatures.}
     \end{center}
\vspace{-0.4cm}
\end{figure}
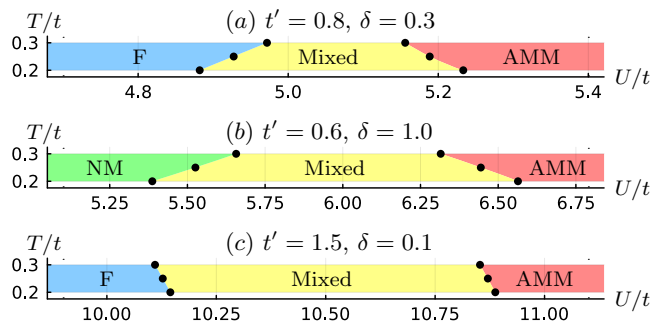

It is well-known that, away from half-filling and at larger $U$-values, ferromagnetic (F) states are relevant for Hubbard-like models \cite{H1963,P1966,N1966}. We therefore also studied solutions with the ansatz $\phi_i=\phiF$, $B_i=\DF\neq 0$ describing F states and, to our surprise, we found F phases in the half-filled DLKK model; see Figs.~\ref{fig3} and \ref{fig4} for examples. 

We also found that mixed phases of different kinds are ubiquitous in the half-filled DLKK model: see Fig.~\ref{fig2}--\ref{fig4} for examples. In particular, it was easy for us to find parameters where the mixed phases are not destroyed by larger temperatures; for example, if $t'/t=1$, then a mixed phase exists even for  $T/t =0.6$, and there are mixed phases for $T/t=0.3$ also for other values of $t'/t$; see Figs.~\ref{fig3} and \ref{fig4}. 
Thus, if the half-filled DLKK is realized in a cold atom system, we predict that mixed phases can be observed.

\paragraph{Conclusions.---}We showed that the DLKK model \cite{D2024} has mixed phases, which are an indication of exotic states where translational invariance is broken in complicated ways, in large parts of the phase diagram at half-filling. 
This opens up for studying exotic physics of the kind that occurs in high-temperature superconductors in cold atom systems. 

There is a common belief that unstable HF solutions cannot occur at half-filling, and that they are peculiar to 2D. Since we found a counterexample to the former, we decided to check the latter by revisiting the  HF solutions used by Penn to obtain his phase diagram of the 3D Hubbard model \cite{P1966}; we expected to confirm Penn's results since his phase diagram is still regarded as state-of-the-art: for example, it was reprinted in a well-cited review on the metal-insulator transition \cite[Fig.~10]{I1998}. However, to our surprise, we found that mixed phases appear also in the 3D Hubbard model and, in fact,  even in the Hubbard model in one and infinite dimensions 
\cite{LL2024A}. 
Thus, mixed phases in the Hubbard model occur in any dimension. 

HF theory is a textbook method. By not checking the stability of solutions of HF equations, one obtains phase diagrams that are qualitatively wrong. Since mean-field phase diagrams can serve as a guide for studies by more sophisticated methods, it is important to get them right. Thus, as the examples in this paper show, HF theory is in need of an update.
In an accompanying more detailed paper, we present such an update, including the corrected phase diagrams of the Hubbard model in various dimensions \cite{LL2024A}. We hope that this will help to avoid misleading HF results about Hubbard-like models in the future.

\paragraph{Appendix: Energy condition.---}
We recall why checking the stability of HF solutions is important. 

At $T=0$, HF theory amounts to finding the Slater state minimizing the expectation value $\langle H\rangle$ of the Hubbard Hamiltonian. This expectation value is a function of the variational fields $\phi_i$ and $B_i$, and the HF equations in \eqref{HFeqs} are the conditions  $\partial \langle H\rangle/\partial \phi_i=\partial \langle H\rangle/\partial B_i=0$. As we all know from calculus, the latter conditions are necessary but not sufficient to characterize the minimizer: energy maxima, saddle points, and local minima that are not absolute all solve the HF equations but have to be discarded because, by a fundamental principle in quantum physics, we need absolute minima of $\langle H\rangle$. 

At $T>0$, we need to minimize the free energy 
$$
\Omega = \Tr(HW) + T\Tr(W\ln(W))
$$
where $W=\exp(-\HHF/T)/\Tr(\exp(-\HHF/T))$ is the thermal state defined by the HF Hamiltonian in \eqref{HHF}. The arguments for $T=0$ above generalize straightforwardly to $T>0$, with $\Omega$ taking the role of $\langle H\rangle$ (note that  $\lim_{T\downarrow 0}\Omega= \langle H\rangle$); the interested reader can find further mathematical details in \cite{LL2024A,BLS1994,BP1996}. 

\bigskip\noindent
{\bf Acknowledgements.} 
We gratefully acknowledge support from the Swedish Research Council, Grants No. 2023-04726 (EL) and  No.\ 2021-03877 (JL).

\end{document}